\documentclass[12pt]{article}

\usepackage{amsmath,amssymb,graphicx,epic}

\newcommand{\be}{\begin{equation}}
\newcommand{\ee}{\end{equation}}
\newcommand{\beq}{\begin{eqnarray}}
\newcommand{\eeq}{\end{eqnarray}}

\begin{document}
\begin{flushright}
{\small  BRX TH-580} \\
{\small  ULB-TH/06-31} \\
\end{flushright}

\vspace{1cm}

\begin{center}
{\Large \sc A note on spin two fields in curved backgrounds}\\[2cm]

S. Deser${}^{a,b}$ and M. Henneaux${}^{c,d}$

\footnotesize \vspace{1cm}

${}^a${\em Brandeis University, Waltham MA 02454, U.S.A.}

\vspace{.2cm}

${}^b${\em California Institute of Technology, Pasadena CA 91125,
U.S.A.}

\vspace{.2cm}

${}^c${\em Universit\'e Libre de Bruxelles and International
Solvay Institutes,
\\ULB-Campus Plaine
C.P. 231\\ Boulevard du Triomphe, B-1050 Bruxelles, Belgium}\\

\vspace{.2cm}

${}^d${\em Centro de Estudios
Cient\'{\i}ficos (CECS), Casilla 1469, Valdivia, Chile}\\

\vspace{2cm}

\begin{tabular}{p{12cm}}
\hspace{5mm}{\bf Abstract:} We reconsider the consistency
constraints on a free massless symmetric, rank 2, tensor field in a
background and confirm that they uniquely require it to be the
linear deviation about (cosmological) Einstein gravity. Neither
adding non-minimal higher derivative terms nor changing the gauge
transformations by allowing terms non-analytic in the cosmological
constant alters this fact.
\end{tabular}
\end{center}

\break

Higher ($s>1$) spin fields are well-known to encounter consistency
problems in curved backgrounds.  This is especially manifest for
massless systems (at least in finite numbers).  The borderline
case is spin 2, where the consistency constraints involve only the
Ricci -- rather than the full Riemann -- tensor
\cite{Aragone:1979bm}. Our note intends to fill a minor gap in the
(correct) belief that a free spin-2 field in a background
describes small excitations off Einstein gravity. We show that the
constraints found in previous treatments cannot be alleviated even
by adding nonminimal terms or by exploiting an apparent additional
freedom in gauge transformations involving terms non-analytic in
the cosmological constant.

We follow the notation of \cite{Deser1}, where details and
conventions may be found.  The action describing a (for notational
convenience only) contravariant tensor density field $h^{\mu \nu}$
in a metric background is \begin{equation} I_2[h] =  \int d^4 x
h^{\mu \nu} \theta_{\mu \nu \alpha \beta} h^{\alpha \beta}
\label{action}
\end{equation} where $\theta$ is the appropriate second order
hermitian operator (generalizing that in a flat background) that
yields the field equation  \begin{equation} 2 G^L_{\mu \nu}(h)
\equiv \Box h_{\mu \nu} - (D^\lambda D_\nu h_{\mu \lambda} +
D^\lambda D_\mu h_{\nu \lambda}) + g_{\mu \nu} D^\alpha D^\beta
h_{\alpha \beta} = 0 \label{FE}
\end{equation} Here all operators, including covariant derivatives
$D_\mu$, are with respect to the background metric $g_{\mu \nu}$.
Choosing a different ordering of derivatives in $\theta$ would lead
to non minimal coupling terms $\sim R h$ in $G_{\mu \nu}^L$. There
is in any case no ordering that preserves the Bianchi identity
$D^\mu G_{\mu \nu}^L = 0$ of flat space. In terms of the action
(\ref{action}), the deviation from the Bianchi identity is
\begin{eqnarray} \delta I_2[h] &=& - 2 \int \xi^\mu D^\nu G^L_{\mu
\nu} (h) \nonumber \\ &\equiv& \int d^4x \xi^\mu \left[ R_{\mu
\sigma} D_\nu h^{\sigma \nu} + \frac{1}{2} \left(D_\alpha R_{\mu
\beta} + D_\beta R_{\mu \alpha} - D_\mu R_{\alpha \beta}\right)
h^{\alpha \beta} \right] \label{devBianchi}
\end{eqnarray} under the gauge transformation \begin{equation}
(-g)^{-\frac{1}{2}} \delta_0 h_{\mu \nu} = D_\mu \xi_\nu + D_\nu
\xi_\mu - g_{\mu \nu} D^\alpha \xi_\alpha. \label{gauge}
\end{equation}
In other words, consistency  -- i.e., vanishing of
(\ref{devBianchi}) for arbitrary $\xi^\mu$ and $h^{\alpha \beta}$
--  requires that the background be Ricci-flat ($R_{\alpha \beta}
= 0$). This ``obstruction" may be restated as the fact that the
system is the linearized deviation of the dynamical Einstein
contravariant metric  density, defined by
\begin{equation} \tilde{\frak{g}}^{\mu \nu} = \frak{g}^{\mu \nu} +
h^{\mu \nu} \label{Einsteinmetric}
\end{equation} as further explained in \cite{Deser1}.

Adding nonminimal terms $\sim \int d^4 x h R h$ - as would also
result from a different ordering -  does not cure the difficulty.
Indeed, the most general quadratic terms\footnote{The terms
allowed for improving the action must be quadratic in $h$:
allowing terms linear in $h$ would merely amount to reconstructing
the Einstein action expanded about a metric which is not solution
of the Einstein equations, and hence would not help.} that can be
added to the action (with proper background covariance) are
\begin{eqnarray} && I_{NM} = a \int d^4x h^{\mu \nu}
R_{\mu \alpha \nu \beta} h^{\alpha \beta} (-g)^{-\frac{1}{2}}+ b
\int d^4 x h^{\mu \alpha} R_{\alpha \beta} h^\beta_{\; \;
\mu}(-g)^{-\frac{1}{2}} \nonumber \\ &&  + c \int d^4 x h^{\mu
\nu} R_{\mu \nu} h(-g)^{-\frac{1}{2}} + \int d^4x \left( d \; R
h^2 + e \; R h_{\mu \nu}h^{\mu \nu} \right)(-g)^{-\frac{1}{2}}
\label{NM}\end{eqnarray} where $h \equiv h^\alpha_{\; \; \alpha}$.
The term involving the full Riemann tensor only makes matters
worse, leaving Riemann dependence in the field equations and
Bianchi ``identities", which would pick up the term $- 2 a \xi^\nu
R_{\mu \alpha \nu \beta} D^\mu h^{\alpha \beta}$, thereby forcing
flatness. Hence one must take $a = 0$. Similarly, the terms
involving the Ricci tensors leave uncancelled $\xi^\alpha
R_{\alpha \beta} D^\beta h$ unless $c=0$ or $\xi^\mu R_{\alpha
\beta} D_\mu h^{\alpha \beta}$ unless $b$ also vanishes. Finally,
the Ricci scalar terms clearly cannot eliminate the Ricci tensor
in the violation of the Bianchi identities. Accordingly, the terms
in (\ref{devBianchi}) cannot be cancelled by variation of
(\ref{NM}), in agreement with the arguments given in
\cite{Aragone:1979bm}.

The above procedure can be generalized slightly -- but
significantly -- by addition of a ``cosmological deviation" term
\begin{equation} I_C = - \frac{\Lambda}{4} \int d^4 x \left(h_{\mu \nu}
h^{\mu \nu} - \frac{1}{2} h^2\right) (-g)^{-\frac{1}{2}} \, ,
\label{cosmo}
\end{equation} whose variation under (\ref{gauge}) is
\begin{equation} \delta I_C = \Lambda \int d^4x \xi_\mu D_\nu h^{\mu
\nu} \, , \label{variationcosmo}
\end{equation} which in turn shifts the Ricci tensor term in
(\ref{devBianchi}) by the cosmological addition $R_{\mu \nu}
\rightarrow R_{\mu \nu} + \Lambda g_{\mu \nu}$.  Correspondingly,
the $h$-field of (\ref{Einsteinmetric}) is now interpreted as the
perturbation off Einstein gravity with a cosmological constant: This
is what consistency now requires of the background.  Note that our
methodology differs slightly from that of \cite{Deser1}, adding a
dynamical term to the spin 2 field and finding that a change is
induced on the background, rather than embedding it in a
cosmological background ab initio.

The remaining hope then is whether the cosmological term
(\ref{cosmo}) can be used to alter the background constraints.
That is, can we alter the gauge transformations such that the
constraint terms requiring $(R_{\mu \nu} + \Lambda g_{\mu \nu}) =
0$ are removed?  This route will now be seen to be ineffective as
well.

Consider a modification of the gauge change non-analytic in
$\Lambda$, permitted by the cosmological term\footnote{This use of
non-analytic terms to restore consistency is borrowed from the
Russian school (e.g., \cite{Fradkin:1986ka} and subsequent work)
where it is used for infinite towers of spins.},
\begin{equation} \delta h_{\mu \nu} = \delta_0 h_{\mu \nu} +
\delta_1 h_{\mu \nu}, \; \;  \delta_1 h_{\mu \nu}= (1/\Lambda)
\Theta_{\mu \nu}^{\; \; \; \; \alpha} \xi_\alpha \label{newgauge}
\end{equation} where $\Theta_{\mu
\nu}^{\; \; \; \; \alpha}$ is an operator of the form $R_{\mu \nu}
D^\alpha$ or $D^\alpha(R_{\mu \nu} \; \; \; \; \; )$.  The idea is
to take advantage of the $\Lambda h_{\mu \nu}^2$ term
(\ref{cosmo}) in the action by adjusting the $1/\Lambda$ term in
(\ref{newgauge}) to cancel the constraint (\ref{devBianchi}). This
can indeed be done, but leaves two residues: the first are $(1/
\Lambda) O(R^2)$ terms from $\delta_1 I_2$, which can (perhaps) be
removed in turn by an iterative procedure, $\delta_2 h_{\mu \nu}
\sim (1/\Lambda^2) O(R^2)$, etc. However, it fails for the very
simple reason that nothing removes the variation $\delta_0 I_C
\sim \int d^4 x \xi_\nu (D_\mu h^{\mu \nu})$ of the cosmological
term, which is the first term in the expansion of the action.  A
$\Lambda^2$-term in the action would be needed, but (purely on
dimensional grounds) there is no local candidate.

One may reformulate the above results in cohomological terms. The
BRST structure for a free spin-2 field in Minkowski space has been
investigated in \cite{Boulanger}. It can then be shown that the
deformation of the model corresponding to a change in the
background is consistent, i.e., defines a cohomological class of
the BRST differential at ghost number zero, only if the modified
background is also a solution of the Einstein equations.

We conclude that there is indeed but one consistent spin-2 model,
and so incidentally only one graviton.

\vspace{1cm} \noindent {\bf Acknowledgements:}  We thank the
organizers of the ``General Relativity Trimester: Gravitational
Waves, Relativistic Astrophysics and Cosmology" at the Institut
Henri Poincar\'e for their hospitality. SD was supported by NSF
grant PHY04-01667. The work of MH is partially supported by IISN -
Belgium (convention 4.4505.86), by the `Interuniversity Attraction
Poles Programme -- Belgian Science Policy' and by the European
Commission programme MRTN-CT-2004-005104, in which he is
associated to V.U. Brussel.

\end{document}